\documentstyle[preprint,aps]{revtex}

\begin{document}
\input epsf
\title{Coherent Atomic Oscillations and Resonances  between
Coupled Bose-Einstein
Condensates with Time-Dependent Trapping Potential}
\author{F.Kh.\ Abdullaev} 
\address{Physical-Technical Institute, Uzbek Academy of Sciences,
\\ 700084, Tashkent-84, G.Mavlyanov str.,2-b, Uzbekistan\\ and\\
Instituto de F\'{\i}sica Teorica, Universidade Estadual Paulista,\\
Rua Pamplona 145, 01405-900, S\~ao Paulo, Brasil }
\author{R.A.\ Kraenkel}
\address{Instituto de F\'{\i}sica Teorica, Universidade Estadual Paulista,\\
Rua Pamplona 145,01405-900, S\~ao Paulo, Brasil\\}
\maketitle

\begin{abstract}
We study the quantum coherent-tunneling between two Bose-Einstein
condensates separated through an oscillating trap potential. The cases of slowly 
and 
rapidly varying in time trap potential are considered. In the case of 
a slowly varying trap we study the nonlinear resonances and chaos in
the oscillations of the relative atomic population. Using the Melnikov
function approach, we find the conditions for  chaotic macroscopic 
quantum-tunneling phenomena to exists. Criteria for the onset of chaos
are also given. We find the values of frequency and
modulation amplitude which lead to chaos on oscillations in the relative 
population, for any given damping and the nonlinear atomic
interaction.  In the case of a rapidly varying trap
we use the multiscale expansion method in  the parameter $\varepsilon =
1/\Omega$, where  $\Omega $ is the frequency of modulations and we derive the
averaged system of
equations for the modes. The analysis of this system  shows that new macroscopic
quantum self trapping  regions, in comparison with the constant trap case,
exist.
\end{abstract} 

\newpage
\section{Introduction}

It has recently been shown that there exists a macroscopic quantum phase 
difference in processes connected with 
atomic waves. Namely, this
effect is observed between two
tunnel-coupled Bose-Einstein condensates \cite{Andr}-\cite{Dalf}. 
The coupling occurs due to a trap potential which has the form of a double-well 
potential.
The theoretical investigation shows that here is possibly an interesting
phenomenon of periodic oscillations of the atomic population between
condensates and quantum self-trapping of population in the dependence of
the relative phase between condensates \cite{Shenoy1}-\cite{Shenoy3}.
Analogous phenomena which have been studied are the ac Josephson
effect \cite{Barone} and the periodic 
exchange of power and switching of electromagnetic waves between cores in 
nonlinear optical couplers \cite{Jensen}.

In particular, there is a direct analogy between the tunneling phenomena  in
two prolongated Bose-Einstein condensates and two tunnel-coupled single
mode optical fibers. 
In this optical analogy, the role of the chemical potential is played the
propagation constant, and to the nonlinear interaction between atoms is 
analogous to
the Kerr nonlinearity of optical media. The tunnnel-coupling arising from
the overlaps of the electromagnetic fields outside of dielectric
cilinders (fibers)\cite{Marcuse,Wab,Abd} exactly corresponds to the tunnel
coupling between two
Bose-Einstein condensates due to the overlaps of the wavefunctions. 

In this context , it is natural to investigate the influence of a time varying
coupling on the quantum coherent tunneling process. In the nonlinear
optical coupler analogy this corresponds to the variation in the
longitudinal direction coupling \cite{Darmanyan,Malomed}.  As was shown
recently, the
variation
in time of the trap potential can lead to  resonant oscillations of the
Bose-Einstein condensate \cite{Castin,Garsia}. The numerical solution of
the 
Gross-Pitaevsky equation shows that the  rapid
variation in time of the trap potential can lead  to a bifurcation 
of the effective (averaged ) 
form of the trap potential and consequently to splitting of
the BEC \cite{Dum}. 

From what we have said above,  we could wait for  new phenomena in the process 
of quantum coherent
atomic tunneling (QCAT) process too. Recalling the analogy with Josephson 
effects in coupled
junctions, we note that in the last case  it is very difficult to produce  such 
kind of effects, the difficulty 
coming from the fact that it
is hard to implement the time varying overlap properties of junctions.

Thus, in this article we will study the atomic tunneling between two
tunnel-coupled BEC's in a double-well {\it time dependent } trap. 
Namely, in Section II we will study the nonlinear resonance phenomena in the 
interference and, in
particular, the stationary regimes, with damping taked into account. The
phase and population damping are considered as well. The phase locked
states are interesting because the suppression in these states of the
fluctuations of the relative phase between two condensates. The regions of
regular and chaotic oscillations of the relative atomic population are
calculated in Section III, using the Melnikov function approach. In 
Section IV the
macroscopic quantum interference in two overlapped condensates with
rapidly varying traps is investigated. We derive the system of averaged
equations for coupled modes and define in Section V new regimes of parameters 
for
the macroscopic quamtum self-trapping (MQST).

\section{Formulation of problem. Slowly varying in time trap}

The problem of Bose-Einstein Condensates in a two double-well time 
dependent trap can be
described by the following model
\begin{eqnarray}
ih\frac{\partial \psi_1}{\partial t} = (E_1(t) + \alpha_1 |\psi_1 |^2
)\psi_1
 - K(t)\psi_2 \nonumber \\
ih\frac{\partial \psi_2}{\partial t} = (E_2(t) + \alpha_2 \psi_2 |^2
)\psi_2 - K(t)\psi_1 ,
\label{basic}
\end{eqnarray}
where the parameters 
$E_i, \alpha_i , K(t)$ are defined by the overlaps integrals of the time
dependent Gross-Pitaevsky eigenfunctions as given in  \cite{Shenoy1}. 
$\alpha_i $ are  parameters of 
nonlinear interactions between atoms and $\alpha_i \sim g_0 = 4\pi h^2
a/m$,
$a$ is the atomic scattering length. The attractive case can be obtained
analogously, using symmetries of the equation
This system of equations is valid in
the approximation of a weak link between condensates. Comparison with the 
numerical solution of GPE shows that 
the  system (\ref{basic}) is a 
reasonable approximation for $z \leq 0.5-0.6.$ For the strongly overlapped 
condensates
the
system should be modified. For a periodically varying $K(t) = K_0 +K_1\sin(\Omega t)$ we have 
limitations on the parameters $K_1$ and $\Omega$ for which the two-mode approximation is 
valid. The small deformation of bound states requires $K_1 \ll K_0$. To
avoid the resonances 
of the modulations  with normal modes of the spherical well we should require that the 
difference of energy between the ground state and the first normal mode be larger than the 
modulation energy, i.e. $ \delta E \gg \hbar \Omega$. Taking into account
that $\delta E \sim 
\hbar \omega_0$, \cite{Zapata}, where $\omega_0$  is the harmonic oscillator frequency, we 
find the condition $\Omega \ll \omega_0$. Taking into account that $\omega_0 \gg \omega_L = 
2K_0$ -- the frequency of the linear oscillations of the atomic population, defined by the 
tunneling frequency, we conclude that by a proper choice of parameters we can have the 
resonant case $ \Omega \simeq \omega_L$ or the rapidly varying modulation $ \Omega \gg 
\omega_L$ . 

Introducing new 
variables $u_i = \sqrt{N_i}\exp(i\theta_i ),  z = (N_1 - N_2)/N_T , N_T =
N_1 + N_2 , \psi = \theta_1 -
\theta_2$, where  $N_i$, and  $\theta_i$ are respectively the number of atoms 
and phases in i-th
trap, we get the  following system
\begin{eqnarray}\label{tun}
z_t = -2K(t)\sqrt{1-z^2}sin{\Phi} - \eta \Phi_t,\\
\Phi_t =\nu \Lambda z + \frac{2K(t)z}{\sqrt{1-z^2}}\cos{\Phi} + \Delta
E(t),
 \end{eqnarray}
where $\Lambda = (\alpha_1 + \alpha_2 ) N_T,$ and $ \nu = \pm 1 $ for the 
positive and negative atomic scattering length respectively. In what follows, 
we will consider the case $\nu = 1$, ($a_s > 0$), if not stated otherwise.
We have included in Eq(\ref{tun}) the damping term $ \eta \phi_{t}(t)$. It 
appears if we
take
into account a noncoherent dissipative current of normal-state atoms,
proportional to the
chemical potential difference $\Delta \mu$. For the other type of
overlapping condensate there might be another type of damping. For example, for 
the
two interacting condensates with different hyperfine levels in a single
harmonic trap the damping has the form $\eta z(t) $.  

The Hamiltonian of the unperturbed system ($K(t) =$ const,$ \Delta E =$ const, 
$\eta = 0$) is
\begin{equation}
H = \frac{\Lambda z^2}{2} - 2K\sqrt{1-z^2}\cos(\Phi) + \Delta E z
\quad .
\end{equation}

Let us consider the case of periodic modulations of the tunnel coupling
parameter, when $K(t) = K_0 + K_1 \sin{\Omega t}$ and the time-periodic energy
difference is $\Delta E(t) = \Delta E + \Delta E_1 \sin(\Omega t)$. 
It is useful  to introduce the parameter
$\Delta = \Phi - \Omega t$, which is a slow-varying function of the time
in a comparison with the period  $\Omega$.
The averaged system of equations can be found by averaging over the period
of fast oscillations $2\pi/\Omega$. Doing so, we get the system 
\begin{eqnarray}\label{av}
\bar{z}_t = -K_1 \cos{\bar{\Delta}}\sqrt{1 - \bar{z}^2} - \eta \Delta_t -
\eta \Omega, \\
\bar{\Delta}_t = -\Omega + \Delta E + \Lambda z - \frac{K_1
z}{\sqrt{1-\bar{z}^2}}\sin{\bar{\Delta}}.
\end{eqnarray}

Let us now look for the fixed points of Eq.(\ref{av}). Introduce first the 
notation $\alpha = \Omega - \Delta 
E$. From the system
(\ref{av}) we find the following equation for the fixed points  
\begin{eqnarray}
z_c = \sqrt{1 - \frac{\eta^2 \Omega^2}{K_1^2 \cos{\Delta_c}^2}},\\
\alpha = z_c (\lambda - K_1^2 \frac{\sin{2\Delta_c}}{2\eta\Omega}).
\end{eqnarray}
For small $z_c , z_c^2 << 1$ we can find the explicit solution  
\begin{eqnarray}
\Delta_c = \arccos(\frac{\eta\Omega}{K_1}), \nonumber \\
z_c \approx \frac{\Omega - \Delta E}{\Lambda - \sqrt{K_1^2 - \eta^2
\Omega^2}}.
\end{eqnarray}

The existence of a stationary solution is connected with the
circumstance that the
damping of the oscillation in quantum tunneling is compensated by the
periodic variation of the trap. In reference\cite{Shenoy2} the influence of the
damping on the macroscopic quantum interference  has been investigated
numerically and
degradation of the interference due to damping has been analyzed.
The predicted stationary solutions,  where these effects are
compensated by the variable driving,  are indeed  phase - locked
solutions. Thus we
can think that the fluctuations of the relative phase will be suppressed
in such states. It is an important prediction since  BEC's have phase
fluctuations and it is difficult to maintain the constant relative phase of
condensates. This phenomenon can be useful to develop the phase standard
for the Bose-Einstein condensate- a problem attracting considerable attention 
recently \cite{Dun}.

We can estimate the value of fixed points for an experimental
situation. For example, when $\Omega = 0.7, \eta = 0.1, K_1
= 0.2, \Lambda = 2.5$, we
obtain the fixed points $z_{1c} =  0.2599, z_{2c} = 0.3037$. 
Stability analysis shows that only the first point is stable and corresponds to
the global attractor.
The results of the numerical simulations of the initial system for the
relative population $z(t)$ are presented in
Fig.1. It shows that the agreement between theory and numerical simulations
is good.


The same system of equations also describes the nonlinear Josephson type 
oscillations in the relative population of a driven , two-component
Bose-Einstein condensate. The two component condensates where two
different hyperfine states can be populated and confined in the same trap.
A weak driving field couples two internal states and as result the
oscillations in the relative population will occur \cite{Williams}.  
In this case the damping term has the form $\Gamma = -\eta z $. Repeating
the above mentioned procedure we obtain the averaged system 
\begin{eqnarray}
\bar{z}_t = -K_1\sqrt{1-\bar{z}^2}\cos{\bar{\Delta}} -\eta
\bar{z},\nonumber \\
\bar{\Delta}_t = \Lambda \bar{z} + \Delta E - \Omega  - \frac{K_1
\bar{z}}{\sqrt{1 -
\bar{z}^2}}\sin{\bar{\Delta}}. 
\end{eqnarray}

In the model \cite{Williams}, $\Delta E $ corrsponds to $\mu_2 -
\mu_1 - \kappa$, where $\mu_i$ are chemical potentials and $\kappa$ is
frequency detuning $\kappa = \omega_d - \omega_0$, $\omega_0$ is the
separation in frequency of two hyperfine states, $\omega_0$ is the
frequency of drive. The fixed points are: 
\begin{eqnarray}
\bar{\Delta}_c = \arccos{\left( \frac{\eta z_c}{K_1
\sqrt{1-z_c^2}}\right)}, \nonumber
\\
\Lambda z_c - \alpha - \frac{K_1 z_c}{\sqrt{1-z_c^2}}(1 - \frac{\eta^2 
z_c^2}{K_1^2 (1-z_c^2)})^{1/2}.
\end{eqnarray}
Here $\alpha = \Omega - \Delta E $.
For the case of small $z_c^2 << 1$ we have the estimate for $z_c$
\begin{equation}
z_{c} = \frac{\Delta E - \Omega}{\Lambda - K_1} + \frac{(\Delta E -
\Omega)^3 (K_1^2 - \eta^2)}{2K_1 (\Lambda - K_1)^4}.
\end{equation}

The typical values of parameters are:$\Omega = 0.7,\mu_2 -\mu_1 - \kappa =
\Delta E = 1, \eta =
0.1, K_1 = 0.2 , \Lambda = 4$, and  the fixed points are $z_{3c} =
-0.111, z_{4c} = -0.13$. The analysis shows that the first fixed point is
stable and second is unstable.

In Fig.2. we present the  plot of solution of the full system for the
damping term
$\eta z$ with the above mentioned set of parameters. All solutions decay
to the first fixed point.

\section{Chaotic oscillations in relative atomic population}

In this section  we consider the interference in the system which in some
limit is mathematically equivalent to the motion of a particle in  a 
double-well potential
under parametric periodic perturbations. The  particle' motion is of
course
unharmonic and the nonlinear resonances between oscillations of a particle
and oscillations of the trap potential are possible. One of consequences of 
this is  the
possible appearence of  chaotic dynamics in such systems. 
We will now study the conditions for the appearence of the
 chaotic dynamics in atomic tunneling. 

To proceed it will be useful first to consider some aspects of the unperturbed
system, that is, when the coupling is constant.  In this case  the system 
(\ref{tun}) is equivalent to the
quartic
(Duffing) oscillator \cite{Shenoy2,Tsironis}. 

We can write the equation for $z(t)$ as
\begin{equation}
z_{tt} = -\frac{\partial V}{\partial z} \quad ,
\end{equation}
where the potential $V(z)$ is given by
\begin{equation}\label{pot}
V(z) = z^2 (a + b z^2), a= 2K_0^2 -
\frac{\Lambda H}{2}, \ b = \frac{\Lambda^2}{8}.
\end{equation}
and $H $ is the Hamiltonian.
The total energy of the effective particle is
$$E_0 = \frac{z_t^2}{2} + V(z) = 2K^2 -H^2/2 \quad . $$
When $E_0 > 0$ we have the oscillating regime with $<z(t)> = 0$ (nonlinear Rabi 
oscillations). This case
corresponds to the periodic flux of atoms from one BEC to other. When $E_0
< 0$ the motion of effective particle is confined in the one of the wells.
This case corresponds to the localization of atomic population in one of
the condensates - the so called macroscopic quantum self-trapping (MQST).
The case
$E_0 = 0$, i.e. $H= 2K$,  corresponds to the separatrix solution separating these two
regimes. It is interesting to investigate the dynamics of atomic
population, when the inital parameters $z(0), \Phi(0)$ are close to
the separatrix of the unperturbed systsem. From the general theory of
nonlinear driven oscillations \cite{Holmes} we can expect the appearence of
{\it chaotic
macroscopic
tunneling phenomena}. 

The separatrix solutions for the right hand side well(condensate) are:
\begin{eqnarray}
z_s(t) = \sqrt{\frac{a}{b}}\mbox{sech}(\sqrt{2a}t) \quad ,\nonumber \\
 \sin^2 (\Phi_s) = \frac{a^2 \mbox{sech}^2 (\sqrt{2a} t)\tanh^2
(\sqrt{2a}t)}{2K_0^2 (b - a \mbox{sech}^2 (\sqrt{2a}t))} \quad ,  
\end{eqnarray}
where $ H = H_s =2K$.

The motion near the separatrix is very sensitive to the change of the initial
condition and a stochastic layer appeas, leading to the chaotic
dynamics of the solutions of the system (\ref{tun}). The Melnikov function
method
is effective to find the regions of chaotic behavior 
\cite{Holmes,Abd3}. The system (\ref{tun}) may written in the form
${\bf r}_t =
{\bf f} + \epsilon {\bf g}$ and has a hyperbolic fixed point.
According to this method, we need  to calculate the Melnikov
function $M(t_0)$
\begin{equation}\label{M}
M(t_0) = \int_{-\infty}^{\infty} (f_{2s} g_{1s} (t,t_0 )- f_{1s} g_{2s}
(t,t_0))dt.
\end{equation}
As is well known, the existence of  zeros of $M(t_0))$ indicates the
intersection of separatrices and the existence of homoclinical chaos.
Substituting the expressions for $f_i, g_i$ from the system
(\ref{tun}) and calculating
the integrals,  we find the final expressions for the Melnikov function. We 
shall give the results for two possible damping terms.
 
\subsection{The case of  the damping in the form $\eta \Phi_t$}

The integrations in (\ref{M}) are cumbersome, and will be omitted. The final 
result for the
Melnikov function is:
\begin{equation}
M_{1}(t_0) = -K_1 F_1 \cos(\Omega t_0 ) - F_2 \eta \quad ,
\end{equation}
where
\begin{eqnarray}
F_1 = \frac{\pi \Lambda  \Omega^2 }{4K_0 b \sinh(\frac{\pi
\Omega}{2\sqrt{2a}})}, \nonumber\\
F_2 = \frac{2\Lambda \sqrt{2a}}{\sqrt{ab -
a^2}}(\Lambda -2H_s )\mbox{arctan}(\sqrt{\frac{a}{b-a}})  -
\frac{\sqrt{2a}}{16(b-a)b^2}G_1  \quad , 
\end{eqnarray}

where

\begin{eqnarray}
G_1 = [-2b^2(2H_s - \Lambda )(bH_s - 4a\Lambda +
3b\Lambda)]\frac{\mbox{arctan}(a/\sqrt{ab-a^2})}{\sqrt{ab-a^2}} - \nonumber\\ 
2\left[ 4 a^2 \Lambda^{2} - 5 ab \Lambda^{2} + 2 b^2 (4 H_s^2 - 4 H_s\Lambda + 
3 \Lambda^{2} )\right] \quad .
\end{eqnarray}

The simple zero of $M(t_0)$ is absent if the condition
\begin{equation}\label{holmes1}
\eta > \frac{F_1}{F_2},
\end{equation}
is satisfied. In Fig.3 we plot this criterion  in the $(K_1 ,\Omega)$ plane for 
 $\eta = 0.1$  . Regions below the curve
correspond to regular oscillations and above to chaotic oscillations of
$z(t)$. The criterion gives a good lower bound on the region of chaos in
this plane. We verified this lower bound  in ($K_1, \Omega$) plane by
numerical simulations and found a good agreement with the formulae
(\ref{holmes1}).

When the damping coefficient is zero we have the expression for the width
of stochastic layer near the separatrix as
\begin{equation}
D = \frac{\pi \Lambda K_1 \Omega^2}{4K_0 b \sinh(\frac{\pi
\Omega}{2\sqrt{2a}})}.
\end{equation}

In Fig.4 we plot the width of the stochastic layer as a function of the 
frequency
for
$K_1 = 0.2, \Lambda = 9.9, z(0) = 0.6, \Phi(0) = 0$.
The maximum of the width is given by the solution of equation
$$\Omega \tanh(\frac{\pi\Omega}{2\sqrt{2a}}) = \frac{4\sqrt{2a}}{\pi}.$$
For this choice of parameters we have $\Omega_m = 1.89$.
For $\Omega \ge 2 , a\ge 1$ we get the estimate $\Omega_m \approx
4\sqrt{2a}/\pi$.
It can be 
seen
that, for the high frequencies $\Omega >> \sqrt{a}$, the stochastic layer
is
exponentially narrow and motion is regular. The analysis of this case will
be performed in detail in the next section. 
In Fig.5 we plot the typical  oscillations in the relative
atomic population for $\Lambda = 9.9, K_0 = 1, K_1 = 0.2,
z(0) = 0.6, \Phi(0) = 0$ for   $\Omega = 1.5$. It turns 
out that for $\Omega$ between $1.5$ and  $3.22$ we have chaotic oscillations, 
and for small or
large $\Omega $, regular oscillations. In Fig.6 the phase portrait is given. 
In Fig.7 the
influence of the damping on chaos is illustrated. Parameters are the same as
in Fig.5 and $\eta = 0.15$. In this case the oscillations become regular.

\subsection {The case of the damping term in the form $\eta z(t)$.}

The calculation of  the integrals which give the Melnikov function is again 
cumbersome, and one find the following result:

\begin{equation}
M(t_0) = F_1 \cos(\Omega t_0) + \eta F_3,
\end{equation}
where 
\begin{equation}
F_3 = \frac{\Lambda -
2H_s}{\sqrt{2(b-a)}}\mbox{arctan}\left( \sqrt{\frac{a}{b-a}}\right)
+ \frac{3\Lambda \sqrt{a}}{2\sqrt{2}b}. 
\end{equation}

The zeros of $M(t_{0})$ are absent when the condition 
\begin{equation}
\eta > \frac{F_1}{F_3},
\end{equation}
is satisfied.

If we consider for example the case  when $K_1 = 0.2, \Omega = 0.7, \Lambda = 
2.5, z(0) = 0.6   ,
\Phi(0) = \pi $, the damping constant should be $\eta = 0.15$. 
In Figs.(7,8) the chaotic and periodic oscillations of the relative population 
in the dependence of
the frequency for fixed $z(0), \Phi(0)$ are plotted.  

\section{Rapidly varying trap potential}

In the rapidly varying case $K =K(\frac{t}{\epsilon})$ we can apply he averaging 
techniques based on
the multiscale expansions \cite{Cole,Kath}.
Let us  search the solution in the form of the sum of the slowly varying
components $U,V$ and rapidly varying corrections $u_i , v_i $
\begin{eqnarray}
\psi_1(t) = U(t_k ) + \epsilon u_1 (\zeta ) + \epsilon^2 u_2 (\zeta )+
...\nonumber \\
\psi_2(t) = V(t_k ) + \epsilon v_2 (\zeta ) + \epsilon^2 v_2 (\zeta ) + ... 
\end{eqnarray}
 where $\zeta = t/\epsilon, \epsilon << 1, $, $t_k = \epsilon^k t $ are
slow times. The averages  over a period of
rapid oscillations are  $<u_i > = <v_i > = 0$. The functions $u_i , v_i $ are
assumed to be functions of the rapid time $\zeta$ and the slow varying
functions $U , V$. The derivatives can be written in the form
\begin{equation}
\frac{d U}{dt} = \frac{dU}{dt_0} + \epsilon\frac{dU}{dt_1} + ...
\end{equation} 
and analogously for the derivative of $V$.

Below, we will assume $E_1 = E_2 ,$. Then the time dependence in $E(t)$ can 
be removed by a simple field transformation $u(v) = u(v)\exp((i/h)\int
E(t')dt')$.
In the zeroth order in $\epsilon$ we get the equations \begin{eqnarray}
ih (\frac{\partial u_0}{\partial t_0} + \frac{\partial u_1}{\partial
\zeta}) = Eu_0 - K(\zeta)v_0 + \alpha |u_0|^2 u_0 \\
ih(\frac{\partial v_0}{\partial t_0} + \frac{\partial v_1}{\partial
\zeta}) = Ev_0 - K(\zeta)u_0 + \alpha |v_0|^2 v_0
\end{eqnarray}

Requiring the exclusion of the secular in $\zeta$ terms, we find the
corrections $u_1 , v_1$
\begin{eqnarray}
u_1 = \frac{i}{h}v_0 (\mu_1 - <\mu_1 >)\\
v_1 = \frac{i}{h}u_0 (\mu_1 - <\mu_1 >)
\end{eqnarray}

where $\mu_1 = \int_{0}^{\zeta}(K(s) - <K>)ds$
and the system for $u_0,v_0$ becomes:
\begin{eqnarray}
ih\frac{\partial u_0}{\partial v_0} = Eu_0 - <K>v_0 + \alpha|u_0|^2 u_0\\
ih\frac{\partial v_0}{\partial t_0} = Ev_0 - <K>u_0 + \alpha |v_0|^2 v_0
\end{eqnarray}

In the order $\epsilon$ we have
\begin{equation}
\frac{\partial u_0}{\partial t_1} = 0, \ \frac{\partial v_0}{\partial t_1} 
= 0 \quad ,
\end{equation}
and we obtain, for the second order corrections, $u_2, v_2$, the elimination of 
the secular terms, results in:
 
\begin{eqnarray}
u_2 = -\frac{i}{h}(\mu_2 - <\mu_2 >)[\frac{\partial v_0}{\partial t_0} +
\frac{i}{h}v_0 + \frac{i\alpha}{h}[2|u_0|^2 v_0 + u_0^2 v_0^* ]]
\nonumber \\
- [(\mu_1 - <\mu_1>)^2 - 2M]u_0  \quad ,\\
v_2 = -\frac{i}{h}(\mu_2 - <\mu_2 >)[\frac{\partial u_0}{\partial t_0} +
\frac{i}{h}u_0 + \frac{i\alpha}{h}[2|v_0|^2 u_0 + v_0^2 u_0^* ]] \nonumber
\\
-[(\mu_1 - <\mu_1> )^2 - 2M]v_0 \quad ,
\end{eqnarray}
where $$\mu_2 = \int_{0}^{\zeta}(\mu_1 {\zeta} - <\mu_1> )ds, \ M =
(<\mu_1^2 > - <\mu_1 >^2 ) \quad .$$
>From this order we get the equations: 
\begin{eqnarray}
\frac{\partial u_0}{\partial t_2} = -\frac{2iM\alpha}{h^3}[2|v_0|^2 u_0 -
v_0^2 u_0^* ] \quad ,\\
\frac{\partial v_0}{\partial t_2} = -\frac{2i\alpha M}{h^3}[2|u_0|^2 v_0
- u_0^2 v_0^* ] \quad .
\end{eqnarray} 
Finally, we have in the next  order, $O(\epsilon^2)$, the averaged system
\begin{eqnarray}
i\hbar\frac{\partial u_0}{\partial t} - Eu_0 - \alpha |u_0|^2 u_0 =
-<K>v_0 + 2\frac{\epsilon^2 \alpha M}{\hbar^2}[2|v_0|^2 u_0 - v_0^2
u_0^*
]\\
i\hbar\frac{\partial u_0}{\partial t} - Ev_0 - \alpha |v_0|^2 v_0 = -
<K>v_0 +
2\frac{\epsilon^2 \alpha M}{\hbar^2}[2|u_0|^2 v_0 - u_0^2 v_0^* ].
\end{eqnarray}
For  periodic modulations of tunnel coupling we obtain $M =
K_1^2/\Omega^2$. 

This averaged system is the main result of this section.
We see that, in comparison with the constant coupling case,  
new effects appear. First, the cross term appears, corresponding  to a 
change of type  in the
effective nonlinearity in coupled BEC' dynamics.
Second, there appears an   additional {\it nonlinear} phase-sensitive coupling.
This term leads to
new minima in the effective potential picture. In the constant coupling
case, the oscillations of $z(t)$ is described by  the Duffing oscillator. 
Now we have a more complicated nonlinear oscillator, describing more
complicated interference patterns.
 
\section{Analysis of the averaged dynamics}
Using the variables introduced in the section II, we get the system 
\begin{eqnarray}\label{rp}
z_t = -2<K> \sqrt{1 - z^2}\sin(\Phi) - \nu \delta (1 - z^2 )\sin(2\Phi)\\
\Phi_t = \Lambda z + 2\frac{<K>z}{\sqrt{1 - z^2}}\cos(\Phi) - \nu \delta z
\left[ 2 
- \cos(2\Phi )\right] .
\end{eqnarray}

where we have defined $\delta = \epsilon^2 \alpha M/h^2 $.
This system has the following Hamiltonian 
\begin{equation}
H = \frac{\Lambda_1 z^2}{2} - 2<K>\sqrt{1-z^2}\cos{\Phi} -
\frac{\delta}{2}(1-z^2)\cos{2\Phi},
\end{equation}
where $\Lambda_1 = \Lambda - 2\delta$.

When the parameter $\delta = 0$, we get back the system (\ref{tun}). The 
effective particle motion
is that of the double nonrigid pendulum system.


Let us consider the different regimes of oscillations.

\subsection{ Linear regime}

This is the case where $|z| << 1, |\Phi| << 1$.
Then, from the system (\ref{rp}), it follows that 
that $$z_{tt} = -2(<K> + \delta )(\Lambda + 2<K>  - \delta )z.$$
Then the frequency of linear oscillations of the number of atoms is
$$\omega_L^2 = -2(<K> + \delta )(\Lambda + 2<K> - \delta ) \quad .$$
Thus,  the frequency of the linear oscillations  grows with  
$\delta\omega^2 = \omega_L^2 - \omega_0^2 = 2\delta(<K> + \Lambda - \delta)$, in comparison 
with a
nonmodulated case. 

\subsection{$z^2 << 1 $, and  $\Phi $ is not small.}

The phase dynamics is described by the equation:
\begin{eqnarray}
\Phi_{tt} + 2 <K >(\Lambda - 2\delta )\sin{\Phi} +  (\delta \Lambda + 2<K>^2 
- 2\delta^2 )\sin(2\Phi) + \nonumber \\
2\delta <K> \sin(3\Phi) + \frac{\delta^2}{2} \sin(4\Phi) = 0. 
\end{eqnarray}
At derivation of this equation we take into account that 
$\Phi_t \sim z$, so we neglect by terms $\sim z \Phi_t$.
This equation is equivalent to a generalized pendulum motion whose 
effective potential is given 
\begin{equation}
U(\Phi) = -2<K>(\Lambda - 2\delta) - \frac{1}{2}(\Lambda \delta + 2<K>^2 -
2\delta^2 )\cos(2\Phi) - \frac{2\delta <K>}{3}\cos(3\Phi) -
\frac{\delta^2}{8}\cos(4\Phi). 
\end{equation}
The additional minima at $\Phi = \pm n\pi$ correspond to the trapping of
oscillations in states witt locked phase.

\section{Macroscopic quantum self-trapping.}

This phenomenon consists in the self trapping of the atomic number in
one of condensates, depending  on  the initial difference in
atomic number and on  the relative phase of condensates . This is connected 
with the
phase sensitive {\it linear coupling}. The nonlinear guided
wave optics analog of this phenomenon, first predicted
in \cite{Shenoy1}, is the  switching of the power in the dual core
couplers from one core to other, depending on the initial power or the 
phase difference. The additional coupling in the averaged system
(\ref{rp}) is {\it nonlinear} and also phase sensitive. Thus the condition
of localization of atomic population should modified in comparison with the one
considered earlier.

Let us find the fixed points of the averaged system (\ref{rp}). 
The first group is for $\Phi_n = \pi n, n= 0,\pm 1,\pm2,...$
For each $\Phi_n$ there can exist one or three roots  for $z$. One of the  
roots is
$z_1 = 0$, two others are  
\begin{equation}
  z_{s1,2} = \pm \left( 1 - \frac{4<K>^2}{(\Lambda -
\delta)^2}\right)^{1/2}. 
\end{equation}
Note that if $(\Lambda - \delta)<K> < 0$, then 3 roots exists for even $n$ 
and
one root ($z_1 = 0$) exist for odd $n$. For $(\Lambda - \delta)<K> > 0 $,one
has the opposite situation, the critical value is $\Lambda_c > 2<K> +
\delta$. 

One can show, however, linear stability analysis predicts that this set is 
unstable when $\delta > 
\Lambda/2$.

The second group of fixed points is given by:
\begin{equation}
\cos(\Phi) = -\frac{<K>}{\sqrt{1-z^2}}.
\end{equation}

Again, there are two possibilities:

i)$\Lambda = 3\delta$, then all line is stationary;

ii)$\Lambda \neq 3\delta$, then one has the following static point
\begin{equation}
\cos(\Phi) = -\frac{<K>}{\delta}, z= 0.
\end{equation} 
This root assumes that $|<K>/\delta| < 1$.
The linear stability analysis shows that the fixed points are unstable
when $\Lambda > 3\delta$.

Following \cite{Shenoy2}, from the constraint coming from the energy conservation and 
the boundness 
of the tunneling energy, given by 

\begin{equation}
H = 
\frac{\Lambda_1 z^{2}(0)}{2} - 2<<K>>\sqrt{1-z^{2}(0)}\cos{(\Phi(0))} -
\frac{\delta}{2}(1-z^{2}(0))\cos(2\Phi(0)) \ge 2<K> -\frac{\delta}{2}
\quad , 
\end{equation}

we obtain the estimate for  the critical value 
of $\Lambda_c$ when the self-localization occurs. We find that:

\begin{equation}\label{cr}
\Lambda_c = \frac{4<K>(\sqrt{1-z^{2}(0)}\cos{\Phi(0)} +1 )
+\delta((1-z^{2}(0))\cos{2\Phi(0)}-1)}{z^{2}(0)} + 2\delta . 
\end{equation}

The phase portrait is plotted in Fig.(9,10,11) for  initial
phase difference $\Phi(0) = \pi$ and $\Lambda = 2.5$, for differnt $ \delta = 
0.2;0.6;0.8$ The regions
corresponding
to   different dynamics of $z(t)$ with $<z> = 0$ and $<z> \neq 0$ are
shown. The increasing of $\delta$ (i.e. $K_1$/$\Omega$) leads to the 
distortion of the MQST regime, and for $\delta \sim 1$ to the nonlinear Rabi-like 
oscillations.
 The typical oscillations in time  of the relative population  for 
different
values of $\delta$ and for fixed $\Lambda$ are plotted in Fig.9. The 
influence
of rapid modulations leads to the appearance of new minima in the
oscillations. This can be described as a result of the overlap of two double
well potentials. This conclusion is  also confirmed by the
results of numerical simulations of the GP
equation for the {\it single} BEC
under rapidly varying trap potential performed in \cite{Dum}. In this work, it
has been shown that the effect of rapid modulations is the appearance of a
effective double well trap potential.


\section{Conclusion}

In this paper we have studied the new effects coming the time
variation of the trap potential, and from damping, on the nonlinear 
oscillations in
the relative population behavior between two BEC's. For the slowly varying
trap we predict synchronization of oscillations of the trap  with
oscillations 
of the relative
population. We find the fixed points for both types of the damping
terms occuring in the studies of two coupled condensates. Using the
Melnikov approach we study the possibility the appearance of the
chaotic oscillations in the tunneling phenomena between two coupled
Bose-Einstein condensates. We calculate the width of the stochastic
layer in phase space, in which the dynamics is chaotic. We find the
lower bound on the region of chaos in $(K_1,\Omega)$ plane. We obtain the
estimate for the 
damping coefficient when the chaos is suppressed. For the rapidly varying
trap we use the multiscale method and derive the averaged equations 
for coupled modes describing the tunneling phenomena. We find the fixed
points in this case describing the stationary states in two coupled 
BEC's. The expression for the critical value of the
nonlinearity parameter , when the macroscopic quantum self-trapping
occurs is derived. Our results show that there is interest to study the 
contribution of the
quantum tunneling for this driven case and the investigation of
resonances and chaos in oscillations of atomic poulation for the strongly
overlapped
condensates \cite{Ost}. These problems require separate investigation. 

\section{Acknowledgments}

 Authors are grateful to E.N.Tsoy for useful discussions and
comments. This work was partially supported by FAPESP Grant. 

 \begin{figure}[t!]\label{Fig.1}
\centering
\hspace*{-5.5mm}
\leavevmode\epsfysize=8cm \epsfbox{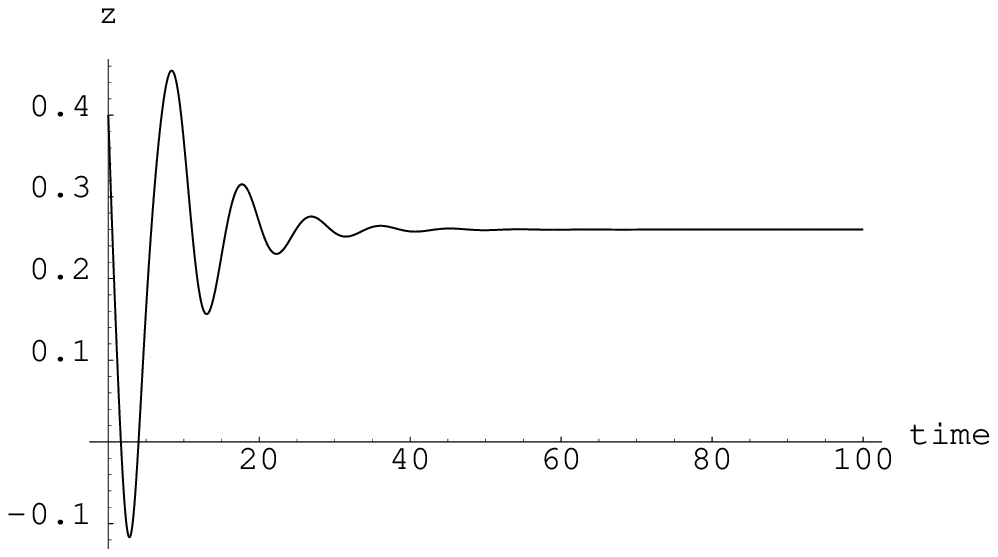}\\[3mm] 
\caption{The typical behavior of the solutions to Eq.(\ref{tun}) when the 
damping is taken in the form $\Gamma = \eta \phi_t$. This
case
corresponds to
$\Omega = 0.7, \eta = 0.1, K_1 = 0.2, \Lambda = 2.5$. Theoretical value
is $z_c = 0.2599$, the numerical is $z_c = 0.26$.}
\end{figure}
\newpage
 
\begin{figure}[t!]\label{Fig.2}
\centering
\hspace*{-5.5mm}
\leavevmode\epsfysize=8cm \epsfbox{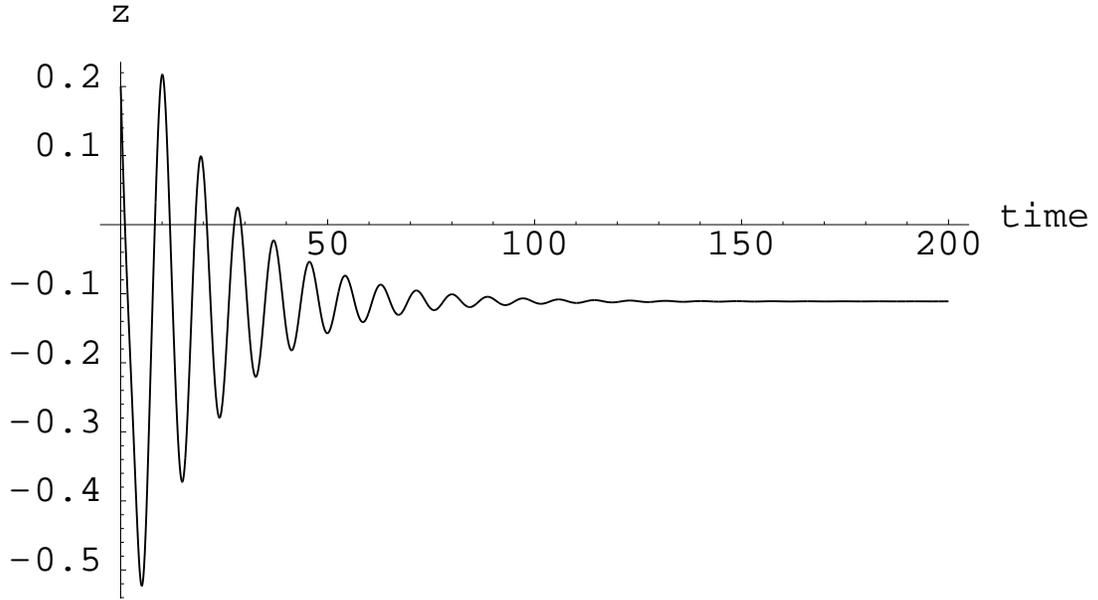}\\[3mm] 
\caption{The typical behavior of the solutions to Eq.(\ref{tun}) when the
damping is taken in the form $\Gamma = \eta z$. This case corresponds to
$\Omega = 0.7, \Delta E = 1, \eta = 0.1, K_1 = 0.2, \Lambda = 2.5$} 
\end{figure}
\newpage

\begin{figure} \label{Fig.3}
 
\hspace{5.5cm}
\leavevmode
\epsfysize=20cm 
\epsffile{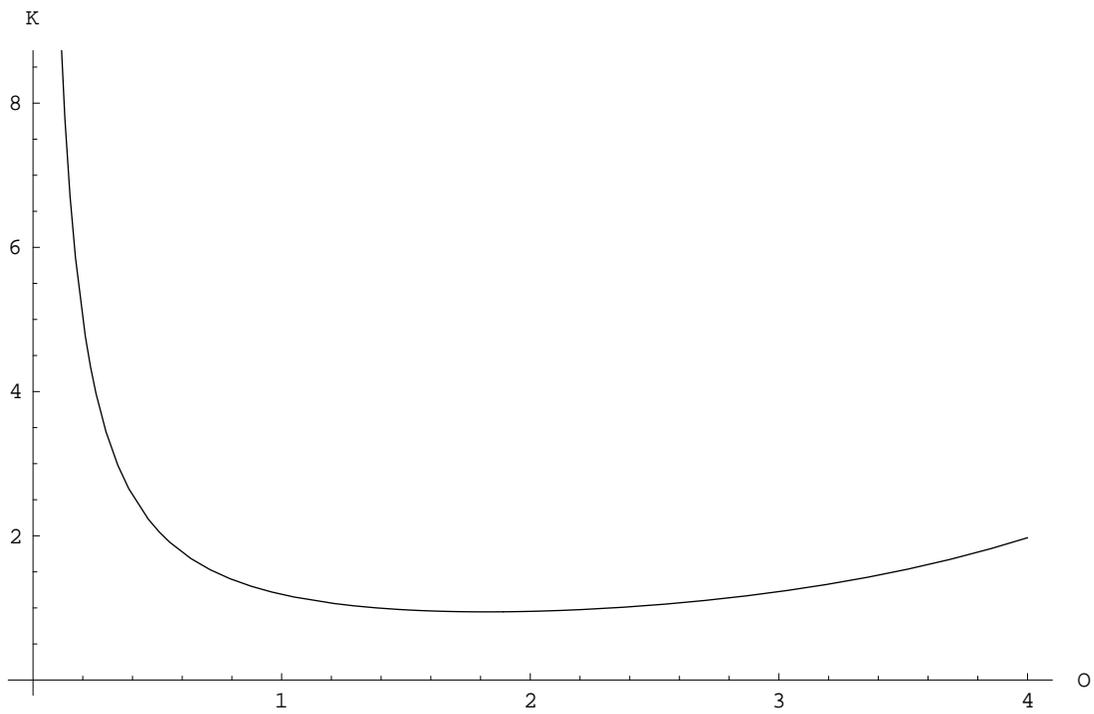}
\caption{The regions of regular and chaotic oscillations in plane
$K_1 , \Omega$ from the Melnikov approach.}
\end{figure}
\newpage
 
\begin{figure}\label{Fig.4}
 
\hspace{5.5mm}
\leavevmode
\epsfysize=20cm 
\epsffile{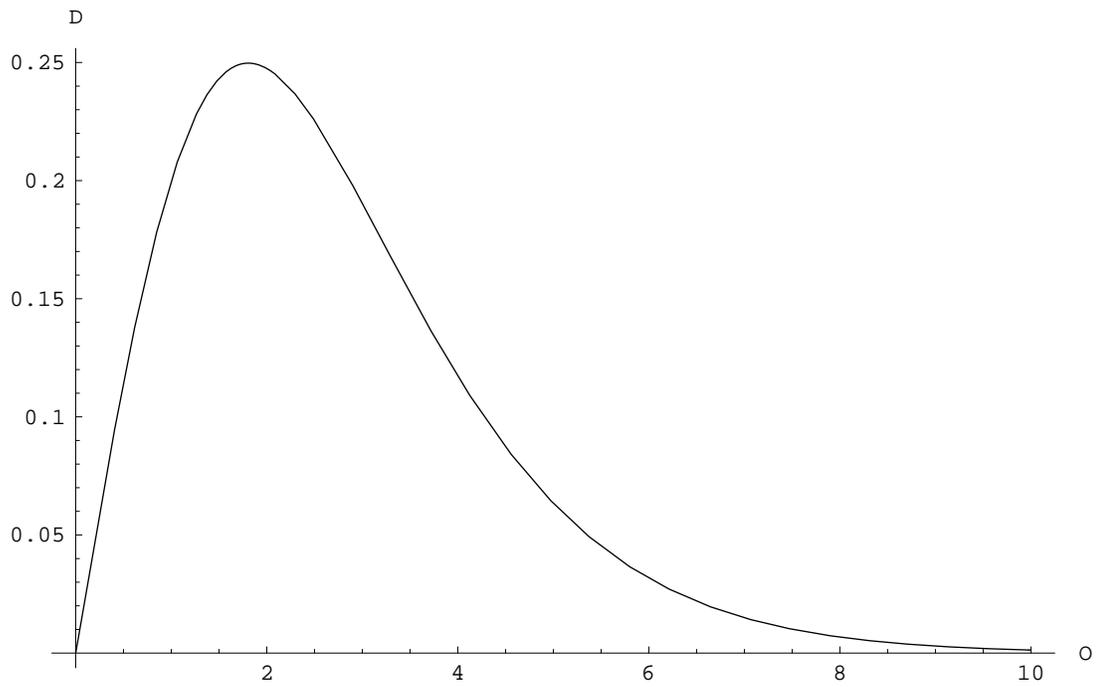}  

\caption{The width of stochastic layer as a function of frequency
$\Omega$,
when $\eta = 0, K_0 = 1, K_1 = 0.2, \Lambda = 9.9, z(0) = 0.6, \Phi(0) = 
0.$} 
\end{figure}
\newpage
 
\begin{figure}\label{Fig.5}
\hspace{5.5mm}
\leavevmode\epsfysize=8cm 
\epsffile{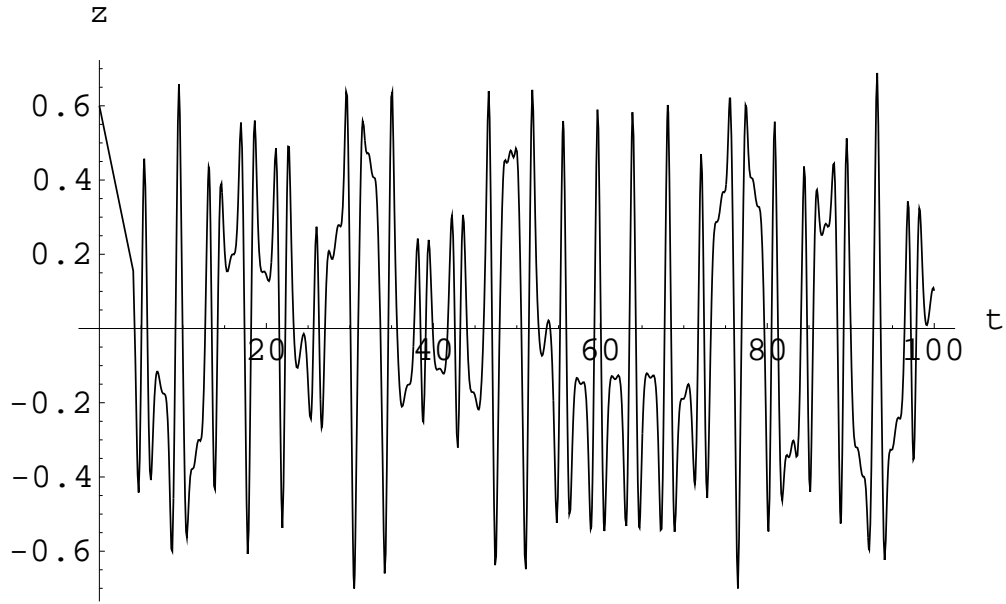}
\vskip 3.0cm
\caption{The typical oscillations of the relative population $z(t)$ for
$  K_1 = 1.2, \Lambda = 9.9, z(0) = 0.6,  \Phi_0 = 0, \Omega = 1.5$}
\end{figure}
\newpage
 
\begin{figure}\label{Fig.6}
\hspace{5.5mm}
\leavevmode\epsfysize=8cm 
\epsffile{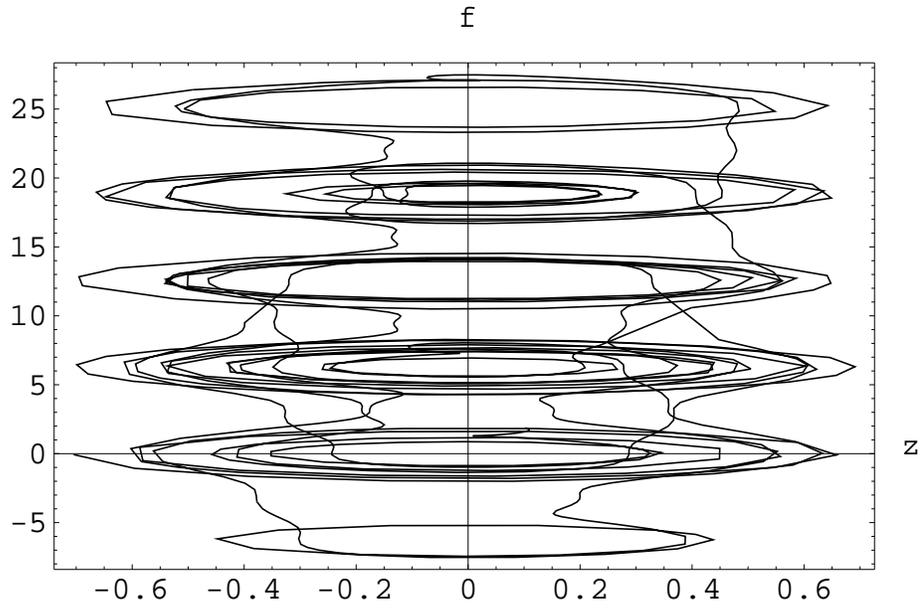}
\vskip 3.0cm
\caption{The  phase portrait in the ($\Phi, z$)-plane for
the same values of the parameters as in Fig(5).} 
\end{figure}
\newpage

\begin{figure}\label{Fig.7b}
 
\hspace{5.5mm}
\leavevmode\epsfysize=8cm 
\epsffile{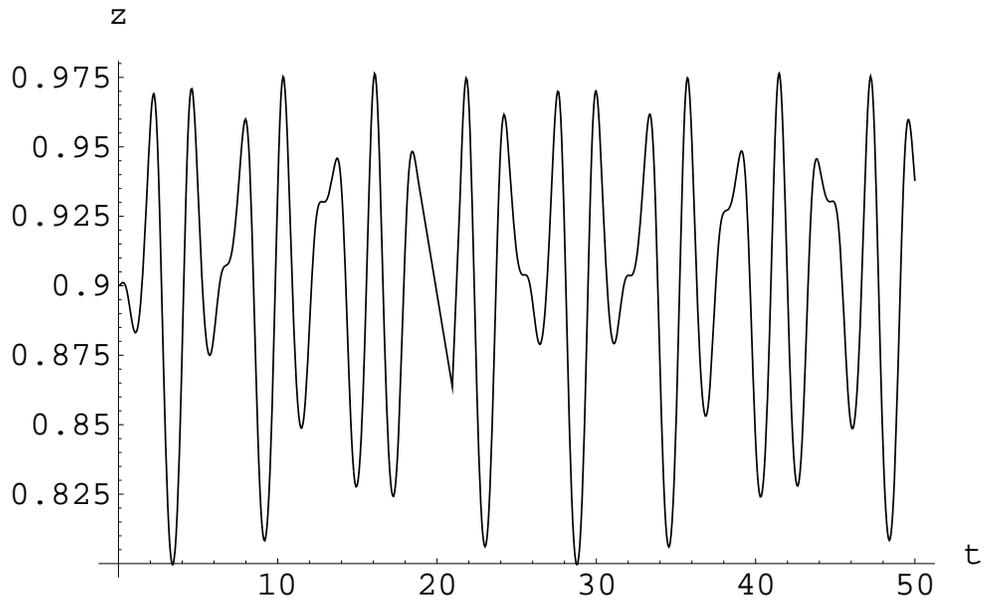}  
\vskip 3.0cm

\caption{ The typical oscillations of the relative population $z(t)$ for $  K_1 
= 0.2, \Lambda = 2.5, z(0) = 0.9,  \Phi_0 = \pi, \Omega = 3.22$} 
\end{figure}
\newpage 

\begin{figure}\label{Fig.7}
 
\hspace{5.5mm}
\leavevmode\epsfysize=8cm 
\epsffile{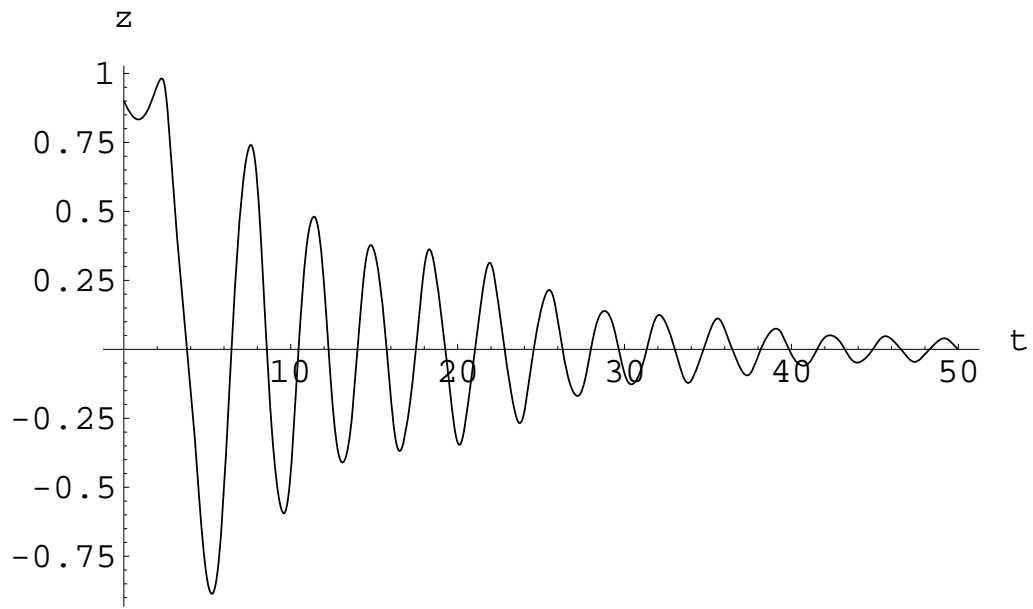}  
\vskip 3.0cm

\caption{The typical oscillations of $z(t)$ for the  same set of
parameters as in Fig.(7), but with  $\Gamma = -0.15 z$}
\end{figure}
\newpage

\begin{figure}\label{Fig.8a}

\hspace{5.5mm}
\leavevmode\epsfysize=8cm 
\epsffile{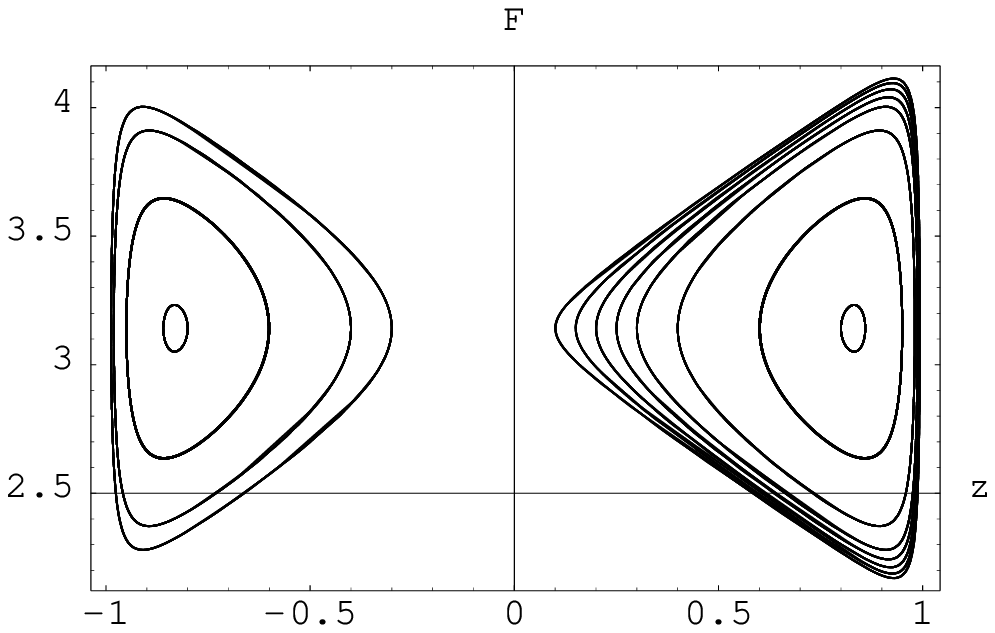}  
\vskip 3.0cm

\caption{The phase portrait in a plane $z, \phi$ of the averaged system
of equations  (\ref{rp}) for $\Phi(0)= \pi, \Lambda = 2.0, \delta = 0.2$}
\end{figure}
\newpage

\begin{figure}\label{Fig.8}

\hspace{5.5mm}
\leavevmode\epsfysize=8cm 
\epsffile{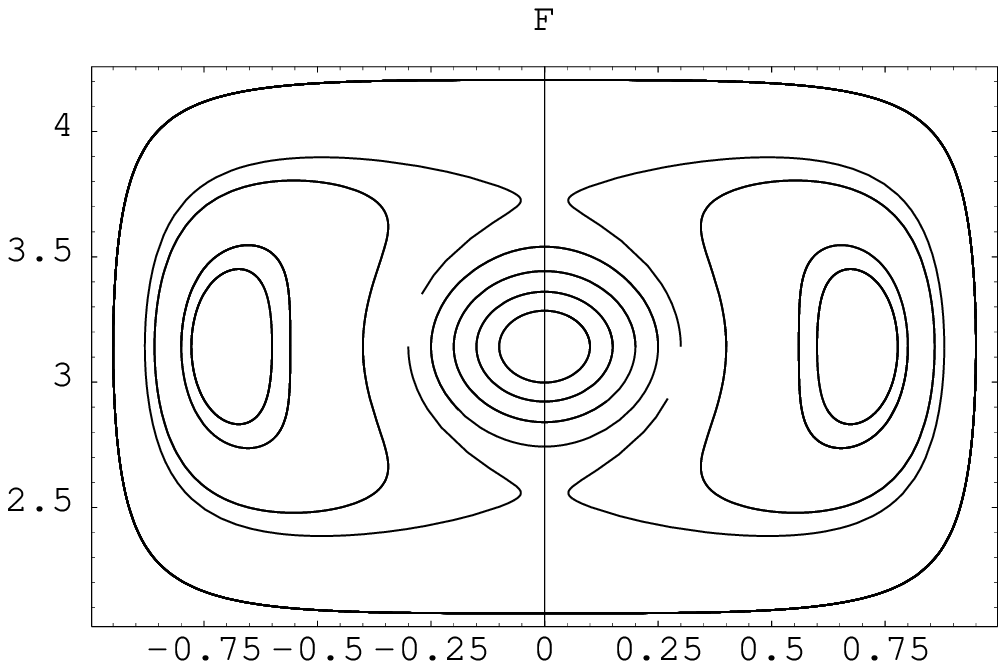}  
\vskip 3.0cm

\caption{The phase portrait in a plane $z, \phi$ of the averaged system
of equations  (\ref{rp}) for $\Phi(0)= \pi, \Lambda = 2.0, \delta = 0.6$}
\end{figure}
\newpage

\begin{figure}\label{Fig.8c}

\hspace{5.5mm}
\leavevmode\epsfysize=8cm 
\epsffile{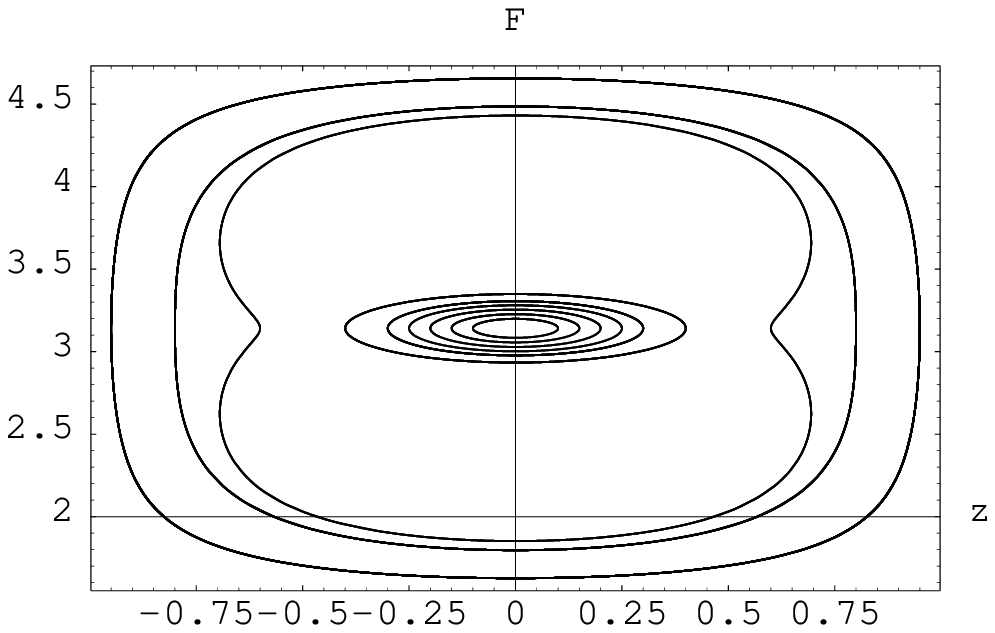}  
\vskip 3.0cm

\caption{The phase portrait in a plane $z, \phi$ of the averaged system
of equations  (\ref{rp}) for $\Phi(0)= \pi, \Lambda = 2.0, \delta = 0.8$} 
\end{figure}
\newpage

\begin{figure}\label{Fig.9}
 
\hspace{5.5mm}
\leavevmode\epsfysize=8cm 
\epsffile{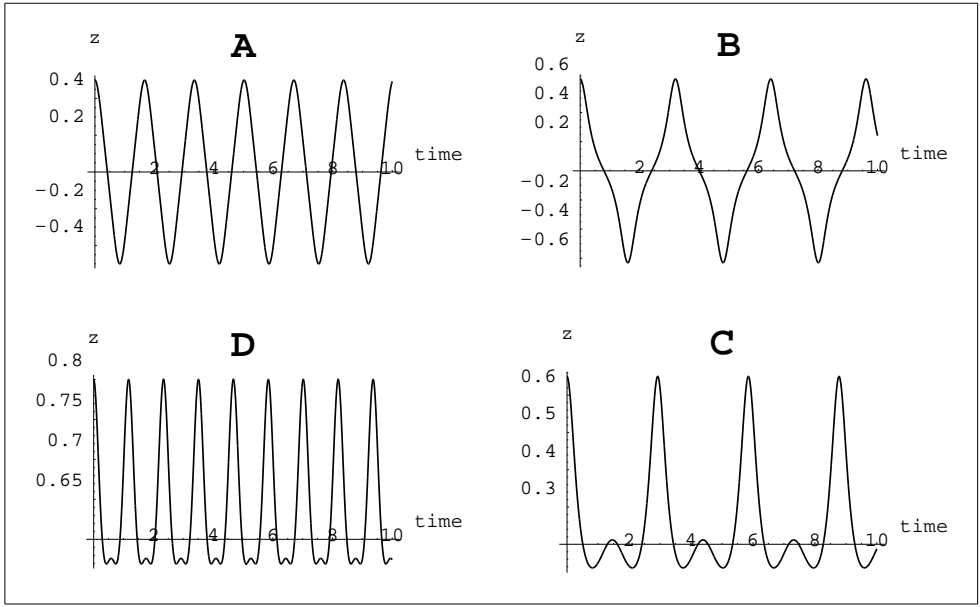}  
\vskip 3.0cm 
\caption{The typical behavior of the solutions to Eq.(\ref{tun}),
for 
$\delta = 0.8$ and $\Lambda = 10,\Phi(0) = 0$ and varying $z(0)$. $({\bf
A})$
corresponds to $z(0) = 0.5$, $({\bf B})$ to $z(0) = 0.65$, $({\bf C})$ to
$z(0) = 0.68$ and ${\bf D}$ to $z(0) =
0.80$. $({\bf B})$ is just before the localization transition and $({\bf
C})$ just after.}
\end{figure}
\newpage

\end{document}